\documentclass[
reprint,
superscriptaddress,
nofootinbib,
amsmath,amssymb,
aps,
]{revtex4-2}
\usepackage{graphicx}% Include figure files
\usepackage{hyperref}
\usepackage{dcolumn}% Align table columns on decimal point
\usepackage{bm}% bold math
\usepackage[mathlines]{lineno}% Enable numbering of text and display math
\usepackage{verbatim}% for multiline comment
\usepackage{hyperref}
\hypersetup{colorlinks,linkcolor={blue},citecolor={blue},urlcolor={blue}}
\begin{document}
\preprint{APS/123-QED}
\title{Nuclear matrix elements calculation for $0\nu\beta\beta$ decay of $^{124}$Sn using nonclosure approach in nuclear shell model}
\author{Shahariar Sarkar}
\email{shahariar.sarkar@iitrpr.ac.in}
\altaffiliation[Present Address: ]{Department of Physics, Indian Institute of Technology Roorkee, Roorkee - 247667, Uttarakhand, India}
\affiliation{Department of Physics, Indian Institute of Technology Ropar, Rupnagar - 140001, Punjab, India}
\author{P. K. Rath}
\affiliation{Department of Physics, University of Lucknow, Lucknow - 226007, India}
\author{V. Nanal}
\affiliation{Department of Nuclear and Atomic Physics,
Tata Institute of Fundamental Research, Mumbai - 400005, India}
\author{R. G.
Pillay}
\affiliation{Department of Physics, Indian Institute of Technology Ropar, Rupnagar - 140001, Punjab, India}
\author{ Pushpendra P. Singh}
\affiliation{Department of Physics, Indian Institute of Technology Ropar, Rupnagar - 140001, Punjab, India}
\author{Y. Iwata}
\affiliation{Osaka University of Economics and Law, Yao, Osaka 581-0853, Japan}
\author{K. Jha}
\affiliation{Department of Physics, Indian Institute of Technology Ropar, Rupnagar - 140001, Punjab, India}
\author{P. K. Raina}
\email{pkraina@iitrpr.ac.in}
\affiliation{Department of Physics, Indian Institute of Technology Ropar, Rupnagar - 140001, Punjab, India}
\date{\today}
%==================================================================================
\begin{abstract}
In this study, we calculate the nuclear matrix elements (NMEs) for the light neutrino-exchange mechanism of neutrinoless double beta $0\nu\beta\beta$) decay of $^{124}$Sn within the framework of the interacting nuclear shell model using the effective shell model Hamiltonian GCN5082. A novel method based on a nonclosure approach is employed, wherein for the intermediate nucleus $^{124}$Sb, effects of energy of 100 states for each $J_{k}^{\pi}$=$0^{+}$ to $11^{+}$ and $2^{-}$ to $9^{-}$ ($\Delta J_{k}$=1) are explicitly included in the NMEs calculation. 
Other common effects such as the finite size of nucleons, higher-order effects of nucleon currents, and short-range correlations (SRC) of nucleons are also taken into account. The extracted optimal closure energy is 2.9 MeV for a total NME of $^{124}$Sn $0\nu\beta\beta$ decay, which is independent of different forms of SRC parametrizations. A comparison of NMEs and half-lives with some of the recent calculations is presented. Further, to gain a comprehensive understanding of the role of nuclear structure on the $0\nu\beta\beta$ decay, the dependence of NMEs on spin-parity of the intermediate states, coupled
spin-parity of neutrons and protons, and the number of intermediate states, is explored.
It is observed that the inclusion of the effects of excitation energies of the intermediate nucleus yields more reliable NMEs. The present findings provide valuable insights for experimental investigations of $0\nu\beta\beta$ decay of $^{124}$Sn
in India and elsewhere. 
\end{abstract}

%\keywords{Suggested keywords}%Use show keys class option if keyword                              %display desired
\maketitle
%\tableofcontents
%=================================================
\section{\label{sec:level1}Introduction}
The neutrinoless double beta ($0\nu\beta\beta$) decay is a rare weak nuclear decay that can occur in certain even-even nuclei. During this process, two neutrons inside the nucleus are converted into two protons and two electrons without emitting any neutrinos. This phenomenon violates the lepton number conservation, and a neutrino is involved as a virtual intermediate particle \cite{Dolinski:2019nrj,Vergados:2016hso,engel2017status,RevModPhys.80.481}.
Observation of this rare decay process would provide strong evidence that neutrinos are Majorana particles. The Majorana nature of neutrinos is a widely favored explanation of the smallness of neutrino mass in many theoretical particle physics models \cite{deppisch2012neutrinoless,PhysRevD.25.2951,rodejohann2011neutrino}. Also, the absolute mass scale of neutrinos is not yet known, and currently, only an upper limit has been derived. The $0\nu\beta\beta$ process is also expected to  provide information on the absolute mass scale of neutrinos \cite{tomoda1991double, RevModPhys.80.481}. The present best upper limit on neutrino mass from $^{136}$Xe $0\nu\beta\beta$ experiment is $\sim 0.160$~eV \cite{PhysRevLett.130.051801}, while that from the tritium beta decay experiment-KATRIN \cite{KATRIN:2021uub} is 0.8 eV. 

Several decay mechanisms have been proposed for the $0\nu\beta\beta$ decay process. These include the widely studied standard light neutrino-exchange mechanism \cite{rodin2006assessment,PhysRevC.60.055502}, the heavy neutrino-exchange mechanism \cite{vergados2012theory}, the left-right symmetric mechanisms \cite{PhysRevLett.44.912,PhysRevLett.47.1713}, and the supersymmetric particle exchange mechanisms \cite{PhysRevD.34.3457,vergados1987neutrinoless}. These mechanisms provide different ways for the decay to occur and are motivated by different theoretical frameworks. In the present work, we focus on the simplest and standard light neutrino-exchange mechanism. 

The decay rate for all $0\nu\beta\beta$ decay mechanisms is related to NMEs and the absolute mass of the neutrino. These NMEs are typically calculated using theoretical nuclear many-body models \cite{engel2017status}. Some of the widely used models are the quasiparticle random phase approximation (QRPA) \cite{vergados2012theory}, the interacting shell-model (ISM) \cite{PhysRevLett.100.052503,PhysRevC.81.024321,PhysRevLett.113.262501}, the interacting boson model (IBM) \cite{PhysRevC.79.044301,PhysRevLett.109.042501}, the generator coordinate method (GCM) \cite{PhysRevLett.105.252503}, the energy density functional (EDF) theory \cite{PhysRevLett.105.252503,PhysRevC.90.054309}, the relativistic energy density functional (REDF) theory \cite{PhysRevC.90.054309,PhysRevC.91.024316}, and the projected Hartree-Fock Bogolibov model (PHFB) \cite{PhysRevC.82.064310}. Recently, some $ab$ $initio$ calculations of NMEs have been performed for the $0\nu\beta\beta$ decay of the lower mass nuclei (A=6-12) using the variational Monte Carlo (VMC) technique \cite{PhysRevC.97.014606,wang2019comparison,PhysRevC.100.055504}. The $ab$ $initio$ methods are highly accurate but computationally intensive, and it is still an open question how well they can be extended to heavier nuclei.

In India, the efforts have been initiated for the TIN.TIN experiment (The INdia’s TIN detector) to search for 0$\nu\beta\beta$ decay in $^{124}$Sn~\cite{nanal2014search}, at
the proposed underground facility of India based Neutrino Observatory (INO) \cite{mondal2012india}. This motivates us to improve the reliability of NMEs for the 0$\nu\beta\beta$ decay of $^{124}$Sn using the nuclear shell model, which will aid in optimizing the experiment setup and extracting the absolute neutrino mass.

The $0\nu\beta\beta$ decay of $^{124}$Sn occurs as 
\begin{equation}
    ^{124}\text{Sn}\rightarrow^{124}\text{Te}+e^-+e^-
\end{equation}
Previously, the nuclear matrix element (NMEs) for the light neutrino-exchange mechanism of 0$\nu\beta\beta$ decay for $^{124}$Sn was calculated using closure approximation in the nuclear shell model in Refs. \cite{PhysRevC.93.024308,Neacsu:2016njp,PhysRevC.98.064324,PhysRevC.98.035502,PhysRevC.101.035504,Menendez:2008jp,PhysRevLett.100.052503}. In the closure approximation, the effects of excitation energy of all the virtual intermediate states of the $0\nu\beta\beta$ decay are approximated with constant closure energy, thereby avoiding the complexity of calculating a large number of intermediate states, which can be computationally challenging for nuclear shell model particularly for higher mass isotopes such as $^{124}$Sn. 
The difficult part of closure approximation is picking the correct closure energy which has no definite method yet and can greatly influence the accuracy of NMEs. 

In recent years, the nonclosure approach has gained popularity due to increasing computational resources. This approach truly includes the real effects of many allowed excitation energy states for each spin-parity of the intermediate nucleus. This way we can improve the reliability of NMEs for $0\nu\beta\beta$ decay and avoid the problem of picking the correct closure energy value. 
This approach was first applied in the nuclear shell model calculations for $0\nu\beta\beta$ decay for $^{48}$Ca \cite{PhysRevC.88.064312}. Subsequently, it was also used for $^{76}$Ge \cite{PhysRevC.93.044334,PhysRevC.90.051301}, and $^{82}$Se \cite{PhysRevC.89.054304}. In two of our earlier studies of $^{48}$Ca \cite{PhysRevC.101.014307,PhysRevC.102.034317}, we have also used this approach. 

In this paper, we  use the nonclosure approach for $^{124}$Sn $0\nu\beta\beta$ decay using the nuclear shell model for the first time. Although it was used extensively for lower mass region isotopes, it is never applied for higher mass region decay candidates of $0\nu\beta\beta$ decay such as $^{124}$Sn. Hence, this study uses the nonclosure approach for the first time to calculate the nuclear matrix elements (NMEs) of $^{124}$Sn, with the aim of examining the effects of excitation energy on a large number of intermediate states. 

This paper is organized as follows. Section \ref{sec:II} outlines the theoretical formalism for computing the nuclear matrix elements (NMEs) for $0\nu\beta\beta$ decay, and presents the expression for the decay rate. In Section \ref{sec:III}, we describe the nonclosure approach to NMEs calculation, which is the method employed in our study. Section \ref{sec:IV} presents the results of our calculations and provides a discussion of the findings, including a comparison to previous studies. Finally, in Section \ref{sec:V}, we summarize the main conclusions of our work.

%=======================================
\section{\label{sec:II}Theoretical formalism of  $0\nu\beta\beta$ decay rate and NME}

%=================
The decay rate for the light neutrino-exchange mechanism of $0\nu\beta\beta$ decay can be written as \cite{PhysRevC.60.055502}
\begin{equation}
    [T^{0\nu}_\frac{1}{2}]^{-1}=G^{0\nu}g_{A}^{4}|M^{0\nu}|^2\left(\frac{\langle m_{\beta\beta}\rangle}{m_e}\right)^2,
\end{equation}
where $G^{0\nu}$ is the well-known phase-space factor that can be calculated accurately \cite{PhysRevC.85.034316}, $M^{0\nu}$ is the total nuclear matrix element for the light neutrino-exchange mechanism, and $\langle m_{\beta\beta}\rangle$ is the effective Majorana neutrino mass defined by the neutrino mass eigenvalues $m_k$ and the neutrino mixing matrix elements $U_{ek}$:
\begin{equation}
\langle m_{\beta\beta}\rangle= \left\lvert\sum_k m_k U_{ek}^2\right\rvert.
\end{equation}
The total nuclear matrix element $M^{0\nu}$ can be expressed as the sum of Gamow-Teller ($M_{GT}$), Fermi ($M_{F}$), and tensor ($M_{T}$) matrix elements, as given by \cite{RevModPhys.80.481}:
\begin{equation}
M^{0\nu}=M_{GT}-\left(\frac{g_V}{g_A}\right)^{2}M_{F}+M_{T},
\end{equation}
where $g_V$ and $g_A$ are the vector and axial-vector constants, respectively. In the present work, $g_V$=1 and the bare value of $g_A$=1.27 is used. The matrix elements $M_{GT}$, $M_{F}$, and $M_{T}$ of the scalar two-body transition operator $O_{12}^\alpha$ of $0\nu\beta\beta$ decay can be expressed as \cite{PhysRevLett.113.262501}:
\begin{eqnarray}
\label{Eq:NMEMAIN}
&&M_{\alpha}=\langle f|O_{12}^\alpha|i\rangle
\end{eqnarray}
where $\alpha\in{F, GT, T}$, and in the present case, $|i\rangle$ corresponds to the $0^+$ ground state of the parent nucleus $^{124}$Sn, and $|f\rangle$ corresponds to the $0^+$ ground state of the granddaughter nucleus $^{124}$Te.

%=================

%==================
The calculation of two-body matrix elements (TBMEs) for $0\nu\beta\beta$ decay involves scalar two-particle transition operators $O_{12}^{\alpha}$ that incorporate both spin and radial neutrino potential operators. These operators are given by \cite{PhysRevC.88.064312}:
\begin{eqnarray}
\label{eq:ncoperator}
O_{12}^{GT}&&=\tau_{1-}\tau_{2-}(\mathbf{\sigma_1.\sigma_2)}H_{GT}(r,E_k),
\nonumber\\
O_{12}^{F}&&=\tau_{1-}\tau_{2-}H_{F}(r,E_k),
\\
O_{12}^{T}&&=\tau_{1-}\tau_{2-}S_{12}H_{T}(r,E_k),
\nonumber
\end{eqnarray}
where, $\tau$ is isospin annihilation operator, $\mathbf{r=r_1-r_2}$ is the inter-nucleon distance of the decaying nucleons, and $r=|\mathbf{r}|$. The operator $S_{12}$ is defined as $S_{12}=3(\mathbf{\sigma_1 .\hat{r})(\sigma_2.\hat{r})-(\sigma_1.\sigma_2)}$.
For the light-neutrino exchange mechanism of $0\nu\beta\beta$ decay, the radial neutrino potential with explicit dependence on the energy of the intermediate states is given by \cite{PhysRevC.88.064312}:
\begin{equation}
\label{eq:npnc}
H_\alpha (r,E_{k})=\frac{2R}{\pi}\int_{0}^{\infty}\frac{h_\alpha(q,r)qdq}{q+E_{k}-(E_{i}+E_{f})/2}, 
\end{equation}
where $R$ is the radius of the parent nucleus, $q$ is the momentum of the virtual Majorana neutrino, $E_{i}$, $E_{k}$ and $E_{f}$ is the energy of initial, intermediate, and final nuclei, and $f_\alpha(q,r)=j_{p}(q,r)h_\alpha(q^2)$ with $j_{p}(q,r)$ is spherical Bessel function ($p=0$ for Fermi and GT, and $p=2$ for tensor NMEs) and $h_\alpha(q^2)$
is the term that accounts for the effects of finite nucleon size (FNS) and higher-order currents (HOC) which are given by \cite{PhysRevC.60.055502,PhysRevC.79.055501}:
\begin{eqnarray}
   h_F(q^2)=&&g_{V}^{2}(q^2),\\
   h_{GT}(q^2)=&&\frac{g_{A}^{2}(q^2)}{g_{A}^2}\left(1-\frac{2}{3}\frac{q^2}{q^2+m_\pi^2}+\frac{1}{3}\left(\frac{q^2}{q^2+m_\pi^2}\right)^2\right)\nonumber\\
    &&+\frac{2}{3}\frac{g_M^2(q^2)}{g_A^2}\frac{q^2}{4m_p^2},\\
   h_{T}(q^2)=&&\frac{g_{A}(q^{2})}{g_{A}^{2}}\left(\frac{2}{3}\frac{q^{2}}{q^{2}+m_{\pi}^{2}}-\frac{1}{3}\left(\frac{q^{2}}{q^{2}+m_{\pi}^{2}}\right)^{2}\right)\nonumber\\
   &&+\frac{1}{3}\frac{g_{M}^{2}(q^{2})}{g_{A}^{2}}\frac{q^{2}}{4m_{p}^{2}}.
   \end{eqnarray}
In this regard, the $g_V(q^2)$, $g_A(q^2)$, and $g_M(q^2)$ form factors, which account for FNS effects, are used. In the dipole approximation, the form factors are given by the following equations \cite{PhysRevC.60.055502,PhysRevC.81.024321}:
 \begin{eqnarray}
        g_V(q^2)=&&\frac{g_V}{\left(1+  \frac{q^2}{M_V^2}\right)^2},\\
     g_A(q^2)=&&\frac{g_A}{\left(1+  \frac{q^2}{M_A^2}\right)^2},\\
     g_M(q^2)=&&(\mu_p-\mu_n)g_V(q^2).
 \end{eqnarray}
where $g_V$, $g_A$, $\mu_p$, and $\mu_n$ are the vector, axial-vector, and magnetic moment coupling constants for the nucleon, and $M_V$, and $M_A$ are the vector and axial-vector meson masses, respectively. The values of $M_V$ and $M_A$ are 850 MeV and 1086 MeV, respectively, while $\mu_p-\mu_n$ is 4.7 is used in the calculation \cite{PhysRevC.81.024321}. The masses of the proton and pion are denoted by $m_p$ and $m_\pi$, respectively.

%========================
In the calculation of the NMEs for $0\nu\beta\beta$ decay, it is also necessary to take into account the effects of SRC. A standard method to include SRC is via a phenomenological Jastrow-like function \cite{PhysRevC.79.055501,vogel2012nuclear}. By including the SRC effect in the Jastrow approach, one can write the NMEs of $0\nu\beta\beta$ defined in Eq. (\ref{Eq:NMEMAIN}) as
 \cite{PhysRevC.79.055501}
\begin{eqnarray}
\label{eq:srcmain}
&&{M}^{0\nu}_{\alpha}=\langle f|f_{Jastrow}(r)O_{12}^\alpha f_{Jastrow}(r)|i\rangle,
\end{eqnarray}
where the Jastrow-type SRC function is defined as 
\begin{equation}
\label{eq:src}
  f_{Jastrow}(r)=1-ce^{-ar^{2}}(1-br^{2}).  
\end{equation}
In literature, three different SRC parametrizations are used: Miller-Spencer, Charge-Dependent Bonn (CD-Bonn), and Argonne V18 (AV18) to parametrize $a, b,$ and $c$ \cite{PhysRevC.81.024321}. The parameters $a$, $b$, and $c$ in different SRC parametrizations are given in Table~\ref{tab:src}. 
%==================
\begin{table}[h!]
\caption{ \label{tab:src}Parameters for the short-range correlation
(SRC) parametrization of Eq. (\ref{eq:src})}.
\begin{ruledtabular}
\begin{tabular}{cccc}
\textbf{SRC Type} & \textbf{a} & \textbf{b} & \textbf{c}\\
\hline
Miller-Spencer&1.10&0.68&1.00\\
CD-Bonn&1.52&1.88&0.46\\
AV18&1.59&1.45&0.92
\end{tabular}
\end{ruledtabular}
\end{table}
%=======================
This approach of using the Jastrow-like function to include the effects of SRC is extensively used in Refs. \cite{PhysRevC.81.024321,neacsu2012fast,Menendez:2008jp}. The authors of Ref. \cite{kortelainen2007short,PhysRevC.75.051303} have recently proposed another method namely: the Unitary Correlation Operator Method (UCOM) \cite{feldmeier1998unitary,NEFF2003311,ROTH20043} to estimate the effects of SRC \cite{vogel2012nuclear}. The present study is beyond the scope of detailed discussion on UCOM approach and we focus only on Jastrow type approach to estimate the effects of SRC. The detailed descriptions of incorporating the SRC effects in different approaches can be found in Refs. \cite{vogel2012nuclear,vsimkovic20090}.

%===================================
\section{\label{sec:III}The Nonclosure Approach of NMEs Calculation}

In the nonclosure approach, the neutrino potential of Eq.(\ref{eq:ncoperator}) is computed explicitly by considering energy $E_k$ of a large number of states $|k\rangle$ of the virtual intermediate nucleus which is $^{124}$Sb for the present case. 

In this approach, the term $E_{k} - (E_{i}+E_{f})/2$ in the denominator of the neutrino potential of Eq. (\ref{eq:npnc}) is written as a function of excitation energy ($E_k^*$) of the intermediate state ($|k\rangle$) as \cite{PhysRevC.88.064312}

\begin{equation}
\label{eq:deno}
E_{k} - \frac{E_i+E_f}{2} \rightarrow \frac{1}{2}Q_{\beta \beta}(0^+)+\triangle M + E_k^*,
\end{equation}
where $Q_{\beta \beta}(0^+)$ is the Q value corresponding to the $0\nu\beta \beta$ decay of $^{124}$Sn, and $\triangle M$ is the mass difference between the $^{124}$Sb and $^{124}$Sn isotopes and $E_k^*$ is the excitation energy of the intermediate states $|k\rangle$ with different allowed spin-parities of $^{124}$Sb.

If one approximates the term $E_{k} - (E_{i}+E_{f})/2$ in the denominator of the neutrino potential of Eq. (\ref{eq:npnc}) with a constant closure energy ($\langle E\rangle$) value such that 
\begin{equation}
    [E_{k}-(E_{i}+E_{f})/2]\rightarrow \langle E\rangle, 
\end{equation}
it is known as closure approximation \cite{PhysRevC.81.024321}. The closure approximation is widely used in the past because it eliminates the complexity of calculating a large number of virtual intermediate states, which can be computationally challenging, particularly for higher mass region isotopes using the nuclear shell model. The difficult part of closure approximation picking the right value of constant closure energy $\langle E\rangle$ which greatly influences the accuracy of the calculated NMEs.  
In this paper, we focus on using the nonclosure approach to include the real effects of at least one hundred states for each spin-parity of the virtual intermediate nucleus $^{124}$Sb. The NMEs with the closure method are also calculated with the closure energy near to the optimal value for which NMEs in closure and nonclosure methods overlap.  

The method based on the nonclosure approach is known as the running nonclosure method \cite{PhysRevC.88.064312} as we can only calculate a finite number of intermediate states out of all possibilities with the current computational limit. The partial NMEs for the transition operator of Eq. (\ref{eq:ncoperator}) in the running nonclosure method can be defined as \cite{PhysRevC.88.064312,PhysRevC.77.045503}.
%========================
\begin{eqnarray}
\label{eq:rncpartial}
&&M_{\alpha}(J_{k},J,E_{k}^{*})=\sum_{k'_{1}k'_{2}k_{1}k_{2}}\sqrt{(2J_{k}+1)(2J_{k}+1)(2J+1)}\nonumber\\
&&\times(-1)^{j_{k1}+j_{k2}+J}
\left\{ \begin{array}{ccc}
j_{k1^{'}} & j_{k1} & J_{k}\\
j_{k2} & j_{k2^{'}} & J
\end{array}\right\}\text{OBTD}(k,f,k'_{2},k_{2},J_{k})\nonumber\\ &&\times \text{OBTD}(k,i,k'_{1},k_{1},J_{k})\langle k_1',k_2':J||\tau_{-1}\tau_{-2}{\mathcal{O}_{12}^\alpha}||k_1,k_2:J\rangle\nonumber\\
\label{eq:mjjkrc}
\end{eqnarray}
%========================

%========================
Here $k_1$ represents a set of spherical quantum numbers ($n_1,l_1,j_1$) for an orbital, similarly for $k_2$, $k_1^{'}$, and $k_2^{'}$. In the present study, $k_{1}$ (and others) has the spherical quantum numbers for $0g_{9/2}$, $1d_{5/2}$, $1d_{3/2}$, $2s_{1/2}$, and $0h_{11/2}$ orbitals for jj55 model space. The $J$ is the allowed spin-parity of the two decaying neutrons and created protons and $J_k$ is the allowed spin-parity of the intermediate states $|k\rangle$. 
The complete expression of non-anti-symmetric reduced TBME ($\langle k_1',k_2':J||\tau_{-1}\tau_{-2}{\mathcal{O}_{12}^\alpha}||k_1,k_2:J\rangle$) for running nonclosure method is given in Ref. \cite{PhysRevC.88.064312}

The one-body transition density (OBTD) are the matrix elements of neutron annihilation and proton creation operators which in proton-neutron formalism can be written as \cite{PhysRevC.88.064312}
\begin{equation}
\label{eq:obtd}
  \text{OBTD}(k,i,k'_{1},k_{1},\mathcal{J})=\frac{\langle k||[a_{k'_{1}}^{+}\otimes\widetilde{a}_{k_{1}}]_\mathcal{J}||i\rangle}{\sqrt{2\mathcal{J}+1}}, 
\end{equation}
where $a_{k'_{1}}^{+}$ and $\widetilde{a}_{k_{1}}$ are the one particle proton creation and neutron annihilation operators, respectively. 
Finally, in the running nonclosure method, the NMEs are calculated by summing over all intermediate states $|k\rangle$ with excitation energies $E_k^*$ up to a certain cutoff value $E_c$ as \cite{PhysRevC.88.064312}
\begin{eqnarray}
{M}_{\alpha}(E_c)=\sum_{J_k,J,E_{k}^{*}\leqslant E_c}{M}_{\alpha}(J_{k},J,E_{k}^{*}).
\end{eqnarray}

The choice of the cutoff energy $E_c$ is important, as it affects the convergence and accuracy of the calculation. Typically, the NMEs are found to be almost constant for values of $E_c$ large enough to include all relevant intermediate states. The most dominating contributions from a few initial states are also observed. 
%========================

%==================
\section{\label{sec:IV}Results and Discussion}
%==================
\begin{table*}
\caption{\label{tab:nmet1} Nuclear matrix elements $M_F$, $M_{GT}$, $M_T$, and $M_{\nu}$ for $0\nu\beta\beta$ decay of $^{124}$Sn, calculated with GCN5082 interaction using running nonclosure and running closure methods for different SRC parametrizations. The closure energy $\langle E\rangle=3.0$ MeV is used for closure NMEs which is near to the optimal value as discussed later.}
\begin{ruledtabular}
\begin{tabular}{cccc}
NME Type&SRC Type&Nonclosure NME&Closure NME\\ \hline
$M_F$&None
&-0.529
&-0.529

\\
$M_F$&Miller-Spencer
&-0.369
&-0.369

\\
$M_F$&CD-Bonn
&-0.564
&-0.565
\\
$M_F$&AV18
&-0.520
&-0.520
\\
\\
$M_{GT}$&None
    &1.961
    &1.954
\\
$M_{GT}$&Miller-Spencer
&1.414
&1.410

\\
$M_{GT}$&CD-Bonn
&2.012
&2.008
\\
$M_{GT}$&AV18
&1.860
&1.854
\\
\\
$M_T$&None
    &0.016
    &0.015
\\
$M_T$&Miller-Spencer
&0.015
&0.014

\\
$M_T$&CD-Bonn
&0.015
&0.014

\\
$M_T$&AV18
&0.015
&0.014
\\
\\
$M^{0\nu}$&None
    &2.304
    &2.297
\\
$M^{0\nu}$&Miller-Spencer
&1.658
&1.654

\\
$M^{0\nu}$&CD-Bonn
&2.377
&2.372

\\
$M^{0\nu}$&AV18
&2.198
&2.192
\\
\end{tabular}
\end{ruledtabular}
\end{table*}
%==================
%==================
\begin{table*}
\caption{\label{tab:nmecomparison} Comparison of NMEs and half-lives of $0\nu\beta\beta$ decay of $^{124}$Sn (the light neutrino-exchange mechanism) calculated with different many-body nuclear models. The phase space factor $G^{0\nu}=9.06\times 10^{-15} (\text{yr}^{-1})$ is used in the current study which is taken from Ref. \cite{PhysRevC.92.055502}. Also, $\left|\eta_{v L}\right|=\frac{\langle m_{\beta\beta}\rangle}{m_e}=10^{-7}$ is used considering upper limit of  $\langle m_{\beta\beta}\rangle$ around 50 meV. The half-lives for other references are recalculated with the quoted NME and $g_{A}$ of those references with the phase space factor and $\left|\eta_{v L}\right|$ of this study.}
\begin{ruledtabular}
\begin{tabular}{ccccccc}
Nuclear Model&Reference&Approximation&$g_{A}$&SRC Type& NME ($M^{0\nu}$)&$\text{T}_{1/2}^{0\nu}$ (yr)\\ \hline
ISM&Current Study&Nonclosure&1.270&CD-Bonn&2.38&$7.49\times 10^{26}$\\

ISM&Ref. \cite{PhysRevC.93.024308}&Closure&1.270&CD-Bonn&2.17&$9.01\times 10^{26}$
\\

ISM&Ref. \cite{Menendez:2008jp}&Closure&1.250&UCOM& 2.62&$6.59\times 10^{26}$
\\

GCM&Ref. \cite{PhysRevC.98.064324}&Closure&1.254&CD-Bonn&2.76 &$5.86\times 10^{26}$ 
\\

IBM-2&Ref. \cite{PhysRevC.91.034304}&Closure&1.269&AV18&3.19&$4.18\times 10^{26}$
\\

QRPA&Ref. \cite{PhysRevC.91.024613}&Closure&1.260&CD-Bonn&5.30&$1.56\times 10^{26}$ 
\\

QRPA&Ref. \cite{PhysRevD.90.096010}&Closure&1.269&AV18&2.56&$6.49\times 10^{26}$
\\

QRPA&Ref. \cite{PhysRevD.90.096010}&Closure&1.269&CD-Bonn&2.91&$5.03\times 10^{26}$
\\

EDF&Ref. \cite{PhysRevLett.105.252503}&Closure&1.250&UCOM&4.81&$1.95\times 10^{26}$
\\

REDF&Ref. \cite{PhysRevC.91.024316}&Closure&1.254&None&4.33&$2.38\times 10^{26}$
\\

\end{tabular}
\end{ruledtabular}
\end{table*}
%==================
The nuclear shell model diagonalization is performed using shell model code KSHELL \cite{Shimizu:2019xcd} to calculate the necessary wave functions and energies of the initial, intermediate, and final nuclei of $0\nu\beta\beta$ decay of $^{124}$Sn. The calculated wave functions are further used to calculate the OBTD that appears in the expression of NME for $0\nu\beta\beta$ decay. The shell model Hamiltonian GCN5082 \cite{PhysRevC.82.064304} of jj55 model space is the used as an input in the calculations. The GCN5082 was also used in the shell model calculations of Ref. \cite{Menendez:2008jp}. Other important Hamiltonian of jj55 model space is SVD which is used in shell model calculations of Ref. \cite{PhysRevC.93.024308}. The SVD Hamiltonian is not publicly available at the time of our calculations, so we are not able to use it in our calculations. For each allowed spin-parity $J_k^{\pi}$ of the virtual intermediate state $^{124}$Sb in the $0\nu\beta\beta$ decay of $^{124}$Sn, we consider the first 100 states. We then calculate the non-anti-symmetric reduced two-body matrix elements for the running nonclosure method using a program that we have written.

%--------------------------------

Table \ref{tab:nmet1} shows the different types of calculated NMEs for $0\nu\beta\beta$ decay of $^{124}$Sn using the nonclosure method in the nuclear shell model. The corresponding results for NMEs in the closure method with closure energy $\langle E\rangle=3.0$ MeV which is near to the optimal value (as discussed later) are also given for comparison. The effects of FNS and HOC are included in all of these NMEs. In addition, results of NMEs for three different parametrizations of SRC (Miller-Spencer, CD-Bonn, and AV18) are also given. It is observed that the GT-type NMEs dominate over the Fermi and tensor-type NMEs, and there is a significant difference in the NMEs depending on the type of SRC used. The SRC type none represents the case where only the effects of FNS and HOC are considered. For Miller-Spencer type SRC, the impact of SRC is most prominent. We see no significant difference in NMEs for closure and nonclosure methods as the closure energy $\langle E\rangle=3.0$ MeV is used which is near to the optimal value for which nonclosure and closure NME overlaps. In the latter part of the discussion, we will describe the method to find the exact optimal closure energy for which closure and nonclosure NMEs overlap. 

In Table \ref{tab:nmecomparison}, we show the newly calculated total NME, calculated with ISM in nonclosure approximation for CD-Bonn SRC and the half-life of $0\nu\beta\beta$ decay of $^{124}$Sn for the light neutrino-exchange mechanism along with the reported results of NMEs and half-lives with different many-body nuclear models, SRC, and approximation. The NMEs vary in the range of 2.15 - 5.30 for various types of nuclear models, approximations and SRC used. In Ref. \cite{PhysRevC.93.024308}, the total NME ($M^{0\nu}$) for the CD-Bonn type of SRC was calculated to be 2.17 with closure energy of 3.5 MeV using the shell model, whereas the NME was reported to be 2.62 in Ref. \cite{Menendez:2008jp}, which was calculated in shell model with UCOM SRC in closure approximation. The newly calculated $M^{0\nu}$ is 2.38 in the present study using the shell model in nonclosure approach for CD-Bonn SRC, which is about 10\% larger as compared to the results of Ref. \cite{PhysRevC.93.024308} and about  10\% smaller as compared to the results of Ref. 
\cite{Menendez:2008jp}. These differences may arise from the choice of Hamiltonian, and closure energy used in the calculations of earlier studies. The NMEs calculated with other nuclear models than the shell model are larger and it's still an open quest to minimize the gaps of this large difference of NMEs in different models. With the newly calculated NMEs, the lower bound of half-life for $0\nu\beta\beta$ decay (light neutrino-exchange mechanism) of $^{124}$Sn is predicted to be $7.49\times 10^{26}$ Years. 
%==========================
%==========================

%---------------------------------
\subsection{Dependence of NMEs on spin-parity ($J_{k}^{\pi}$) of the intermediate states of $^{124}$Sb}
%==================

To study the contribution of each allowed spin-parity state of the virtual intermediate nucleus $^{124}$Sb on the NMEs, we use the expression of Eq. (\ref{eq:rncpartial}) and calculate the partial NMEs as:
\begin{eqnarray}
{M}_{\alpha}(E_c,J_k)=\sum_{J,E_{k}^{}\leqslant E_c}{M}_{\alpha}(J_{k},J,E_{k}^{}).
\end{eqnarray}
Here, $J_k^{\pi}$ represents the spin-parity state of the intermediate nucleus $^{124}$Sb. 
%==================
\begin{figure}[h]
\includegraphics[trim=2.5cm 1cm 2.5cm 2cm,width=\linewidth]{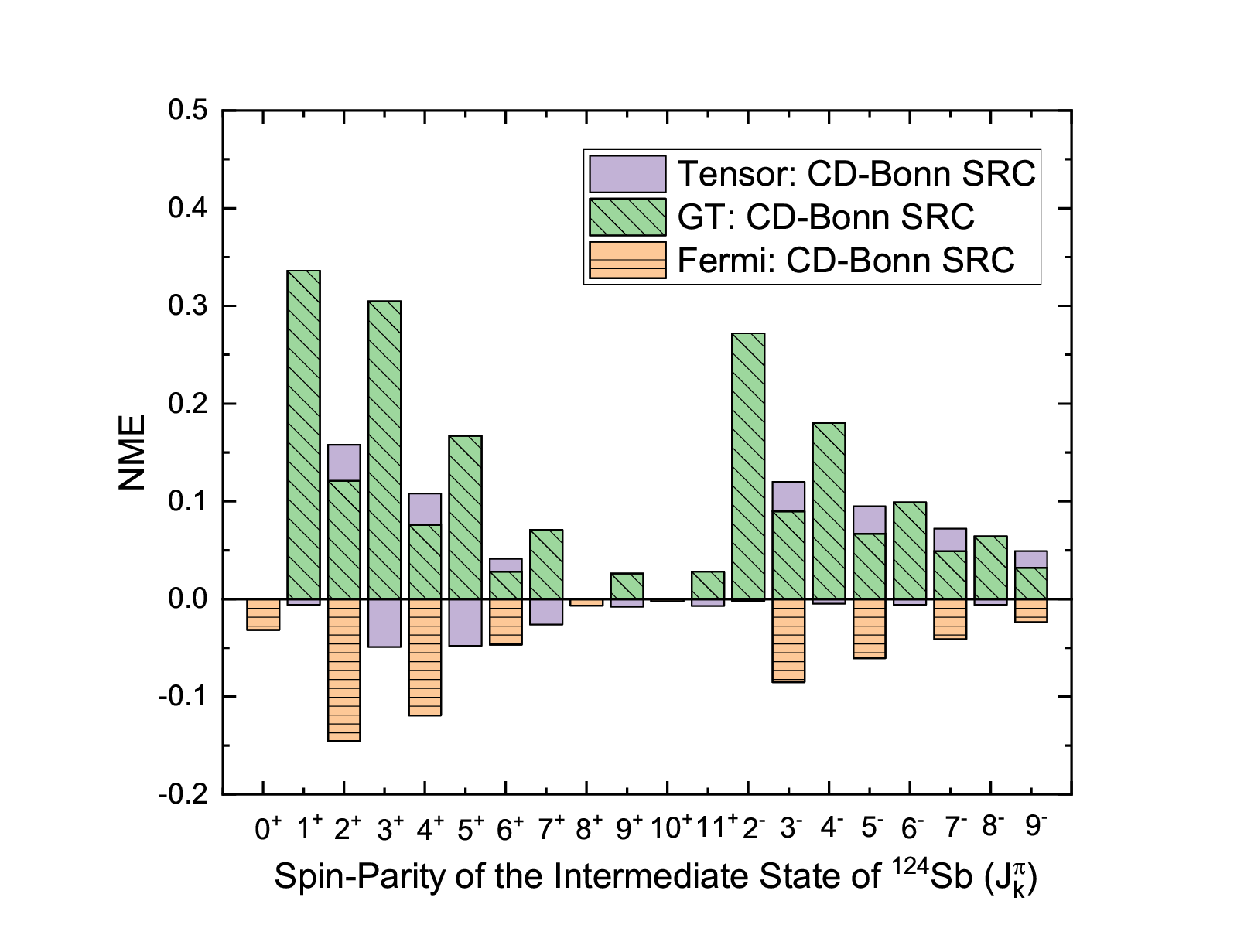}% Here is how to import EPS art
\caption{\label{fig:NMEvsJk}(Color online) The figure shows the contribution through different spin-parity of virtual intermediate states of $^{124}$Sb ($J_{k}^{\pi}$) in NMEs for the light neutrino-exchange mechanism of $0\nu\beta\beta$ decay of $^{124}$Sn. The NMEs are calculated using the running nonclosure method with GCN5082 effective interaction and CD-Bonn SRC parametrization.}
\end{figure}
%=======================

Figure~\ref{fig:NMEvsJk} shows the dependence of the different types of NMEs on the spin-parity states of $^{124}$Sb. We find that for all Fermi-type NMEs, the contribution through each $J_{k}^{\pi}$ is negative, whereas for all Gamow-Teller type NMEs, the contribution is positive. In the case of tensor NMEs, the contributions from different $J_{k}^{\pi}$ states come in the opposite phase, reducing the total tensor NMEs.

We observe that the most dominant contribution to $M_F$ type NMEs comes from the 2$^{+}$ state, with significant contributions from 4$^{+}$, 6$^{+}$, 3$^{-}$, 5$^{-}$, and 7$^{-}$, states. For $M_{GT}$ type NMEs, all $J_{k}^{\pi}$ contribute significantly, except for the 0$^{+}$ state. The most dominant contributions come through 1$^{+}$ state. For $M_{T}$ type NMEs, prominent negative contributions come from 3$^{+}$, 5$^{+}$, and 7$^{+}$ states, with the contributions from 3$^{+}$ state being the most dominant. Contributions from 2$^{+}$, 4$^{+}$, 6$^{+}$, 3$^{-}$, 5$^{-}$, 7$^{-}$, and  9$^{-}$ states are all positive. We observe a similar pattern of variation of different types of NMEs with $J_{k}^{\pi}$ for other types of SRC parametrization as well.

%==================
\subsection{Dependence of NMEs on coupled spin-parity ($J^\pi$) of two decaying neutrons and two created protons}

We have also analyzed the dependence of NMEs on the coupled spin-parity ($J^\pi$) of the two decaying neutrons and two created protons in the decay. To do this, we use the running nonclosure method with the CD-Bonn SRC parametrization and write the NMEs as
\begin{eqnarray}
{M}_{\alpha}(E_c,J)=\sum_{J_k,E_{k}^{}\leqslant E_c}{M}_{\alpha}(J{k},J,E_{k}^{})
\end{eqnarray}

\begin{figure}[h]
\includegraphics[trim=3cm 1cm 2cm 2cm,width=\linewidth]{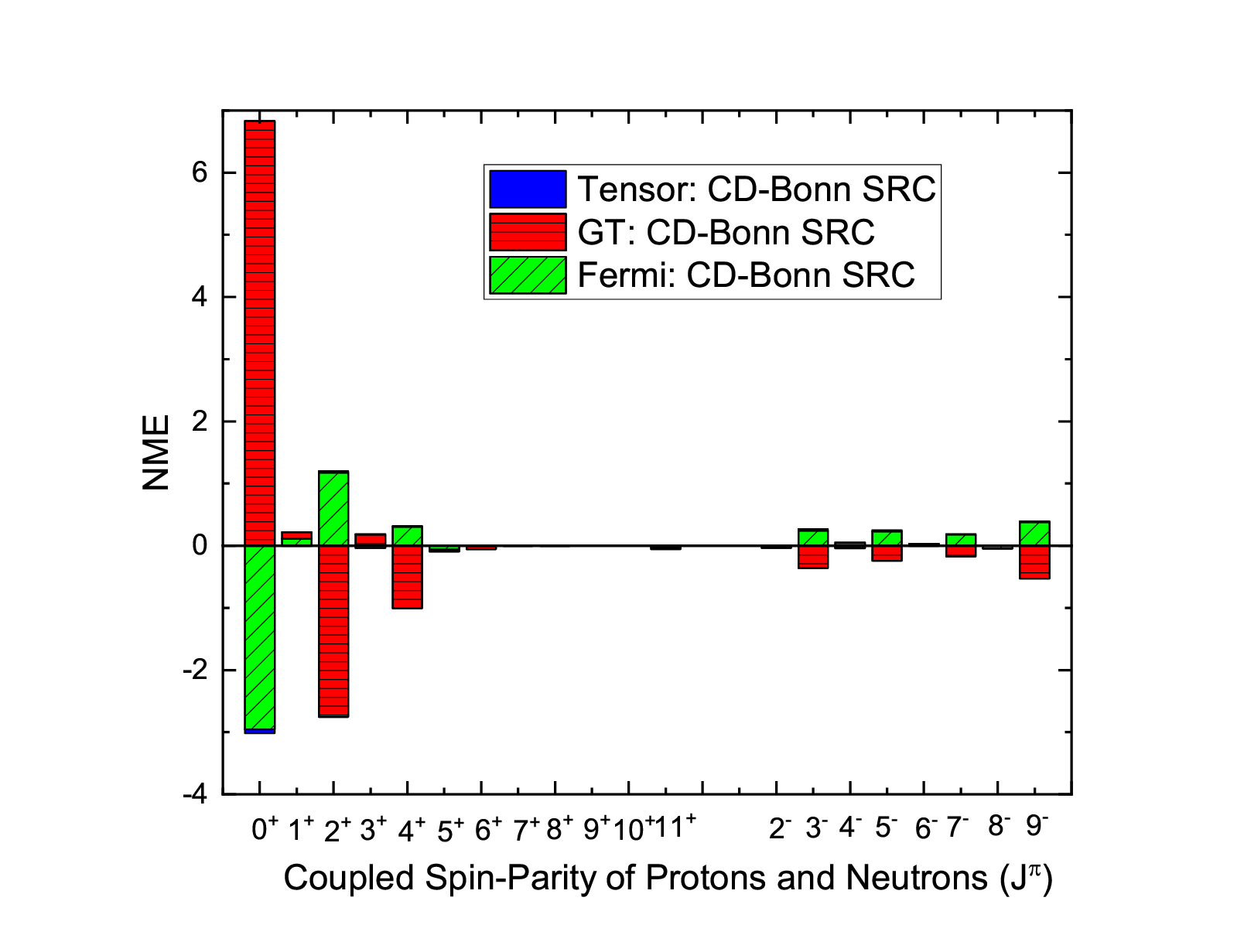}% Here is how to import EPS art
\caption{\label{fig:NMEvsJ}(Color online) Contribution through different coupled spin-parity of two initial neutrons or two final created protons ($J^{\pi}$) in NMEs for the light neutrino-exchange mechanism of $0\nu\beta\beta$ decay of $^{124}$Sn. The results show NMEs calculated in running the nonclosure method with GCN5082 effective interaction for CD-Bonn SRC parametrization.}
\end{figure}

The contributions of NMEs through different $J^\pi$ are shown in Fig.~\ref{fig:NMEvsJ}. 
%==================
%=======================
We observe that for all types of NMEs, the most dominant contributions come from the 0$^{+}$ and 2$^{+}$ states. Additionally, the contribution from 0$^{+}$ and 2$^{+}$ states have opposite signs, leading to a reduction in the total NMEs. There are also small contributions from the 4$^{+}$ and 6$^{+}$ states, with almost negligible contributions from odd-$J^\pi$ states. This is due to the pairing effect, which is responsible for the dominance of even-$J^\pi$ contributions \cite{PhysRevLett.113.262501}.

\subsection{Variation of NMEs for $0\nu\beta\beta$ with the cutoff number of states ($N_c$) of $^{124}$Sb}

%==================
\begin{figure*}
\centering
\includegraphics[trim=0cm 0cm 0cm 0cm,width=\linewidth]{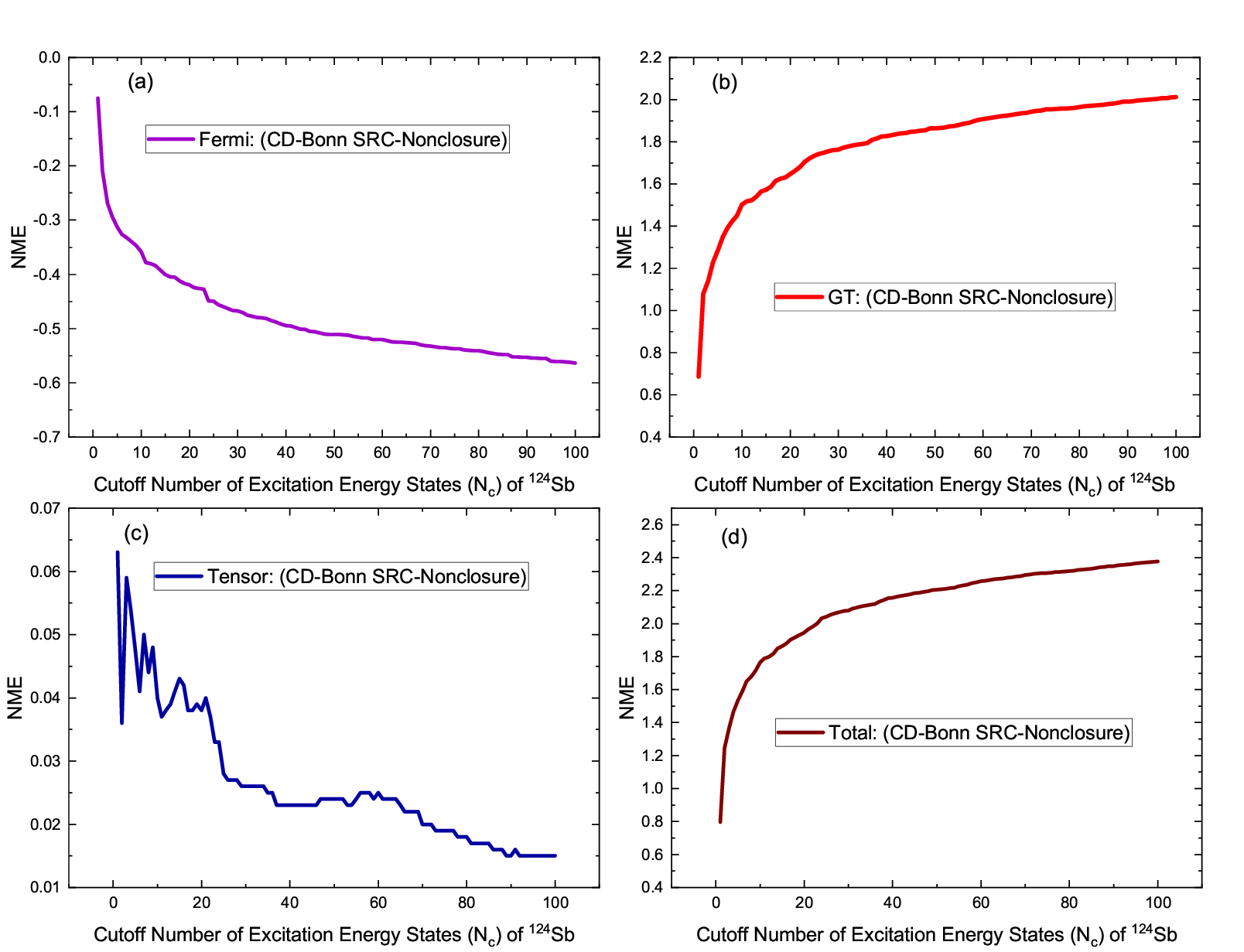}% Here is how to import EPS art
\caption{\label{fig:nmevsnk}(Color online) The figure shows the variation of (a) Fermi, (b) Gamow-Teller, (c) tensor, and (d) total nuclear matrix elements (NMEs) for the $0\nu\beta\beta$ (the light neutrino-exchange mechanism) of $^{124}$Sn with the cutoff number of excitation energy states ($N_c$) of the virtual intermediate nucleus $^{124}$Sb. The NMEs are calculated using the total GCN5082 interaction for the CD-Bonn short-range correlation (SRC) parametrization in the running nonclosure method.}
\end{figure*}
%==================
%==================
\begin{figure*}
\centering
\includegraphics[trim=0cm 0cm 2.5cm 0cm,width=\linewidth]{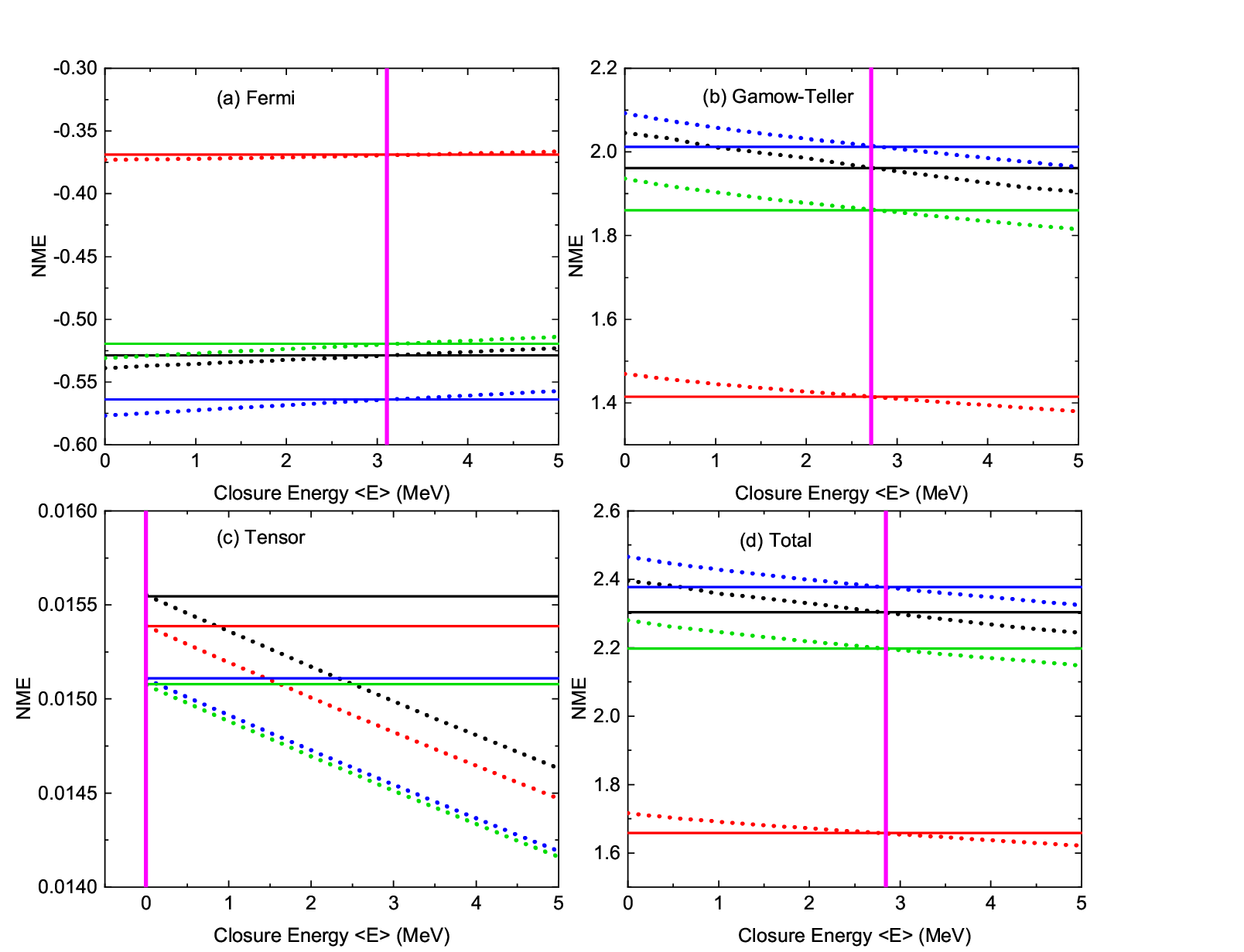}
\caption{\label{fig:optimalclosure}(Color online) The figure shows the dependence of closure NMEs of (a) Fermi (b) Gamow-Teller (c) Tensor, and (d) Total types with closure energy $\langle E\rangle$ for different SRC parametrizations. The plot also shows the nonclosure NMEs to find the crossover of closure and nonclosure NMEs (marked with a vertical magenta line), which is the optimal closure energy for which the closure and nonclosure NMEs overlap. The dotted lines (all colors) represent closure NMEs and the solid lines (all colors) represent nonclosure NMEs. Also, black lines (dotted and solid) represent SRC-none case, whereas, the red, blue, and green lines (dotted and solid) represent NMEs for Miller-Spencer, CD-Bonn, and AV18 type SRC, respectively.}
\end{figure*}
%==================
%==================
To assess the impact of the number of states on the calculated NMEs, we examine the dependence of the NMEs on the cutoff ($N_c$) for each allowed $J_{k}^{\pi}$ of $^{124}$Sb. We express the NMEs as a function of $N_c$ in the running nonclosure method as

\begin{eqnarray}
{M}_{\alpha}(N_c)=\sum_{J_k,J,N_k\leqslant N_c} {M}_{\alpha}(J_{k},J,N_k),
\end{eqnarray}
where ${M}_{\alpha}(J_{k},J,N_k)$ is same as defined in Eq. (\ref{eq:rncpartial}).

The dependence of the different types of NMEs on cutoff number of excitation energy states ($N_c$) of $^{124}$Sb is shown in Fig.~\ref{fig:nmevsnk}. We find that the first few low-lying states contributes the most, but after $N_c=50$, the different types of NMEs reach mostly a stable value. For larger $N_c$, the NMEs become mostly constant. To obtain NMEs with negligible uncertainty, we were able to consider $N_c=100$ for each allowed $J_k^{\pi}$ of $^{124}$Sb with our available computational facility. While the values of the tensor type NMEs are not yet saturated even at $N_c=100$, this is not a problem as the contribution from tensor type NMEs is negligible compared to the GT and Fermi type NMEs. We note that a similar dependence of NMEs on $N_c$ is seen for other SRC parametrizations.

%======================
\subsection{Finding the Optimal Value of Closure Energy for $0\nu\beta\beta$ decay of $^{124}$Sn}
%======================
Here, we discuss the important issue of identifying the optimal closure energy for which the NMEs in the closure and nonclosure approaches overlap. In order to accomplish this, we plot variations of Fermi, GT, tensor, and total NMEs calculated using the closure approach for different SRC parametrization with closure energy as shown in Fig. \ref{fig:optimalclosure}. The nonclosure NMEs are also shown in Fig. \ref{fig:optimalclosure}. At optimal value of closure energy, there is a crossing of the nonclosure and closure NMEs as indicated by a vertical magenta line in Fig. \ref{fig:optimalclosure}.

The optimal closure energy for Fermi-type NMEs is determined to be around 3.1 MeV for various SRC parametrizations, whereas, the optimal closure energy is about 2.7 MeV for GT NMEs. The optimal closure energy for tensor-type NMEs is around 0 MeV. Since the GT component dominates on the total NMEs, therefore, optimal closure energy for total NME is about 2.9 MeV which is similar to GT type NMEs.

In the end, by determining the optimal closure energy, it is easy to obtain the nonclosure NMEs using the closure approximation and avoid the complexity of computing a large number of states in the nuclear shell model. We showed the results of closure NMEs along with nonclosure NMEs in Table \ref{tab:nmet1} with closure energy $\langle E\rangle=3.0$ MeV which is near the optimal value for total NMEs. Hence, closure and nonclosure NMEs in Table \ref{tab:nmet1} are very similar. 
%==================
\subsection{NMEs for $2\nu\beta\beta$ decay and its dependence on the number of states and excitation energy of $1^{+}$ state of $^{124}$Sb}
%==================
\begin{table*}
\caption{\label{tab:nmedbd} The Calculated NME and half-life for $2\nu\beta\beta$ decay of $^{124}$Sn using nuclear shell model. Phase-space factor $G^{2\nu}=5.31\times10^{-19}$ (yr$^{-1}$) is used which is taken from Ref. \cite{PhysRevC.93.024308}. The half-lives for other references are recalculated with the quoted NME and $g_{A}$ of those references with the phase space factor of this study.}
\begin{ruledtabular}
\begin{tabular}{ccccc}
Nuclear Model&Reference&$g_{A}$&NME ($M_{GT}^{2\nu}$ (MeV$^{-1}$)&$\text{T}_{1/2}^{2\nu}$ (yr)\\ \hline
ISM&Current Study&1.270&0.054&$0.95\times 10^{21}$ \\
ISM&Ref. \cite{PhysRevC.93.024308}&1.270&0.042&$1.6\times 10^{21}$\\
ISM&Ref. \cite{caurier1999shell}&1.250&0.101&$0.29\times 10^{21}$\\
QRPA&Ref. \cite{suhonen1998weak}&1.254&0.193&$0.078\times 10^{21}$\\
QRPA&Ref. \cite{suhonen2011double}&1.250&0.110&$0.24\times 10^{21}$\\
QRPA&Ref. \cite{vsimkovic20130}&1.000&0.200&$0.18\times 10^{21}$\\

\end{tabular}
\end{ruledtabular}
\end{table*}
%==================
To conclude our investigation, we finally calculate the NME and half-life for $2\nu\beta\beta$ decay of $^{124}$Sn and examine the dependence of NMEs for two-neutrino double beta ($2\nu\beta\beta$) decay of $^{124}$Sn on the number of states and excitation energy of the virtual intermediate nucleus $^{124}$Sb. The $2\nu\beta\beta$ decay process is similar to $0\nu\beta\beta$ decay, except that $2\nu\beta\beta$ is a lepton number-conserving decay where two anti-neutrinos appear in the final state, along with two electrons. The $2\nu\beta\beta$ decay of $^{124}$Sn to $^{124}$Te, along with two electrons and two anti-neutrinos, is expressed as

\begin{equation}
^{124}\text{Sn}\rightarrow^{124}\text{Te}+e^-+e^-+\overline {\nu}_e+\overline {\nu}_e.
\end{equation}

The half-life of the $2\nu\beta\beta$ decay of the $0^+$ ground state to the $0^+$ ground state transition is given by \cite{Doi:1985dx,Tomoda:1990rs,Haxton:1984ggj}

\begin{equation}
[T^{2\nu}_\frac{1}{2}]^{-1}=G^{2\nu}g_{A}^{4}|m_{e}c^{2}M_{GT}^{2\nu}|^{2},
\end{equation}
where $G^{2\nu}$ is the phase-space factor \cite{Doi:1985dx,PhysRevC.75.034303}. In this case, only the Gamow-Teller type NMEs ($M_{GT}^{2\nu}$) are relevant, and they can be written as \cite{Doi:1985dx,PhysRevC.75.034303}

\begin{equation}
M_{GT}^{2\nu}=\sum_{k,E_{k}^{}\leqslant E_c}\frac{\langle f||\sigma\tau_{2}^-||k\rangle\langle k||\sigma\tau_{1}^-||i\rangle}{E_k^{*}+E_0},
\end{equation}
where $\tau^-$ is the isospin lowering operator. In this study, $|i\rangle$ represents the $0^+$ ground state of the parent nucleus $^{124}$Sn, $|f\rangle$ represents the $0^+$ ground state of the grand-daughter nucleus $^{124}$Te, and $|k\rangle$ represents the $1^+$ states of the intermediate nucleus $^{124}$Sb. $E_k^{*}$ is the excitation energy of the $1^{+}$ states of $^{124}$Sb, and the constant $E_0$ is given by

\begin{equation}
E_0=\frac{1}{2}Q_{\beta \beta}(0^+)+\triangle M. 
\end{equation}
Here $Q_{\beta \beta}(0^+)$ is the Q value corresponding to the $\beta \beta$ decay of $^{124}$Sn, and $\triangle M$ is the mass difference between the $^{124}$Sb and $^{124}$Sn isotopes. For the calculation, the bare value of $g_{A}=1.27$ is used.

%==================
\begin{figure}[h]
\centering
\includegraphics[trim=2cm 1cm 2cm 2cm,width=\linewidth]{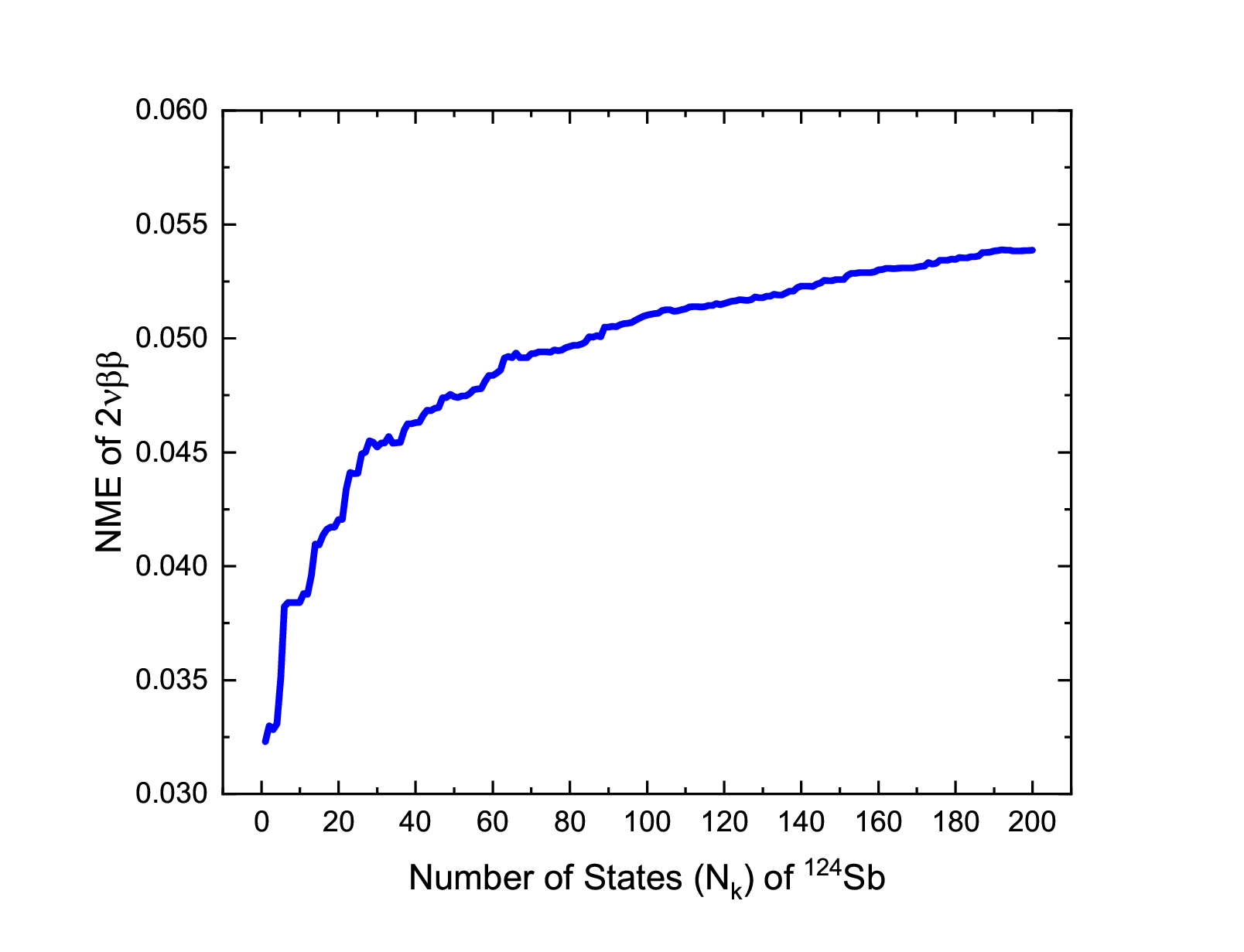}% Here is how to import EPS art
\caption{\label{fig:nmevsnumdbd} Variation of the NMEs for the $2\nu\beta\beta$ decay of $^{124}$Sn with the number of $1^{+}$ states ($N_k$) of the virtual intermediate nucleus $^{124}$Sb.}
\end{figure}
%==================
%==================
\begin{figure}
\centering
\includegraphics[trim=2cm 1cm 2cm 2cm,width=\linewidth]{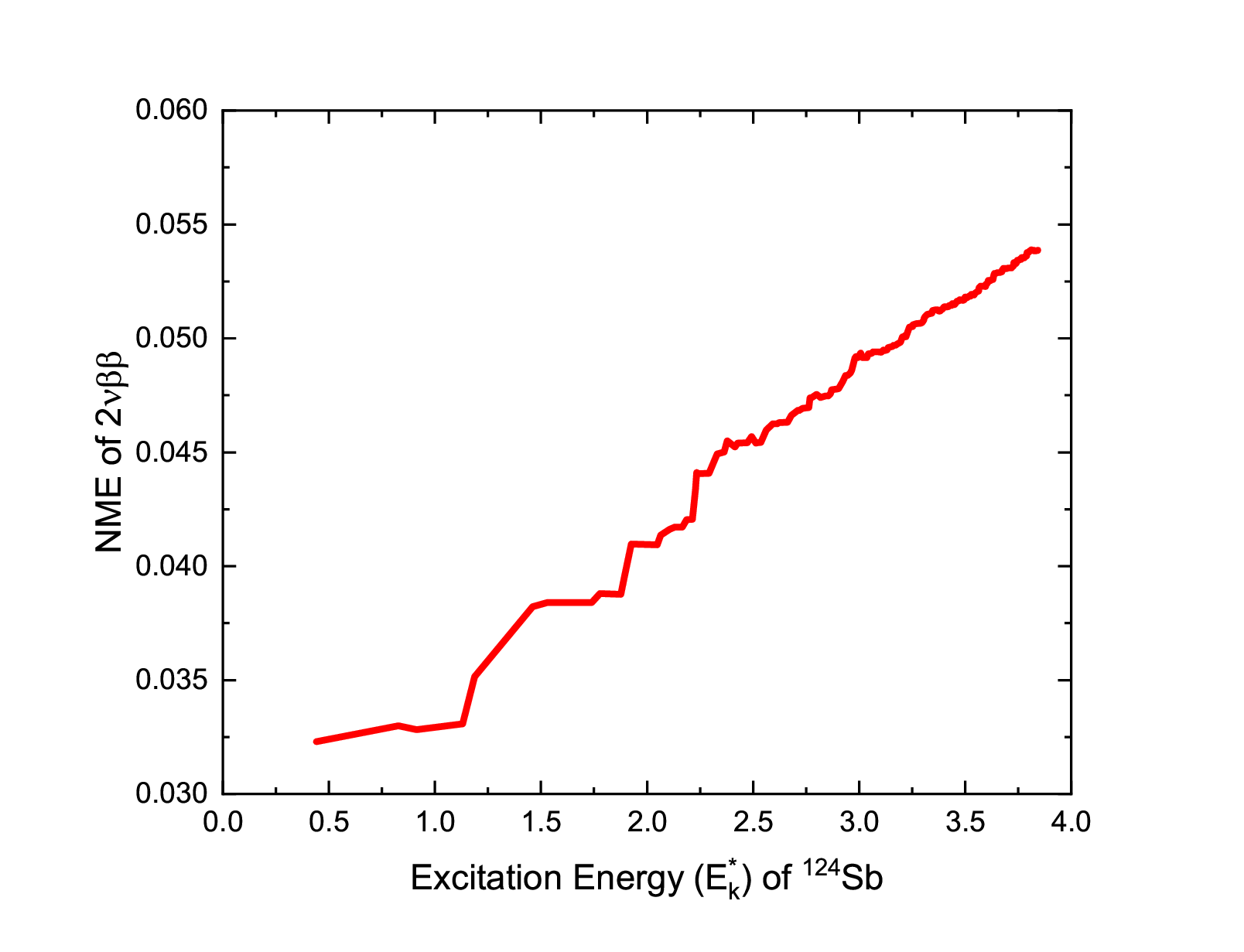}% Here is how to import EPS art
\caption{\label{fig:nmevsexcdbd} The variation of the NMEs for the $2\nu\beta\beta$ decay of $^{124}$Sn with excitation energy ($E_{k}^{*}$) of $1^{+}$ states in the virtual intermediate nucleus $^{124}$Sb is shown.}
\end{figure}
%===========================
The calculated NME for $2\nu\beta\beta$ decay of $^{124}$Sn is given in Table \ref{tab:nmedbd}. The calculated NME is 0.054 when the first two-hundreds $1^{+}$ states of the virtual intermediate nucleus $^{124}$Sb are considered. With the calculated NME, the half-life for $2\nu\beta\beta$ decay of $^{124}$Sn is predicted to be $0.95\times 10^{21}$ (yr). In this Table, values of NMEs and half-lives of $2\nu\beta\beta$ decay of $^{124}$Sn from some of the recent calculations are also given. 

Figure \ref{fig:nmevsnumdbd} illustrates the variation of the NMEs for $2\nu\beta\beta$ decay, computed using the total GCN5082 interaction, with the number of $1^{+}$ states ($N_k$) considered for the virtual intermediate nucleus $^{124}$Sb. Our calculations with the available computational facility were able to account for the effects of the first 200 $1^{+}$ states of $^{124}$Sb, which resulted in a mostly saturated value of NMEs. As $N_k$ increases, the NMEs reach a constant value, indicating that further inclusion of $1^{+}$ states beyond this point will not significantly affect the computed value.

In Fig. \ref{fig:nmevsexcdbd}, we demonstrate the dependence of NMEs for $2\nu\beta\beta$ decay on the excitation energy ($E_{k}^{*}$) of $1^{+}$ states in the virtual intermediate nucleus $^{124}$Sb. Our computations, which incorporated the effects of the first 200 $1^{+}$ states of $^{124}$Sb using the KSHELL software, were able to capture excitation energies up to about 4 MeV, resulting in computed NMEs as shown in Fig.  \ref{fig:nmevsexcdbd}. We anticipate that considering more $1^{+}$ states, up to around 10 MeV, will lead to a fully saturated NMEs value. While our current computational resources allowed us to incorporate 200 $1^{+}$ states in months, we aim to use enhanced computing facilities and allocate more time in future calculations to include more states.

%==================
\section{\label{sec:V}Summary and Conclusions}
The study of $^{124}$Sn is of great significance as it is an important candidate for $0\nu\beta\beta$ with immense interest from the double beta decay community of India. The NMEs play a vital role in extracting the neutrino mass from the measured half-life of $0\nu\beta\beta$ decay. %making them a crucial focus for the experimental neutrinoless double beta decay community in India. 

In the present work, we have employed the nonclosure approach to calculate the NMEs with improved reliability for $0\nu\beta\beta$ decay of $^{124}$Sn (a light neutrino-exchange mechanism) in the nuclear shell model, which explicitly takes into consideration of excitation energy for one hundred states of each spin-parity of the intermediate nucleus $^{124}$Sb.
%The aim was to improve the reliability of the NMEs which is the central theme of studying $0\nu\beta\beta$ decay. 
It is observed that the present method resulted in a 10\% variation in NMEs, as compared to the recent closure approach calculations using a nuclear shell model with different input Hamiltonian and closure energy. Hence, the difference may arise either due to the choice of the input Hamiltonian or the choice of closure energy in the earlier studies. %, which is very difficult to pick without knowing the nonclosure NMEs.
With the present values of NMEs, the lower bound on the half-life of $0\nu\beta\beta$ decay (a light neutrino-exchange mechanism) of $^{124}$Sn is predicted to be $7.49\times 10^{26}$ Years.

Furthermore, we have analyzed the dependence of NMEs on the spin-parity of intermediate states in $^{124}$Sb, as well as that of coupled protons and neutrons. The contributions of each spin-parity of the intermediate states were all positive for GT-type NMEs and negative for Fermi-type NMEs. For coupled spin-parity of protons and neutrons, the $0^{+}$ and $2^{+}$ contributed the most in all NMEs. 
The effect of the number of intermediate states on the saturation of NMEs has also been investigated. The NMEs mostly attained a constant value when one hundred states for each spin-parity of the intermediate nucleus $^{124}$Sb was considered.  

It may be pointed out that the choice of the closure energy is arbitrary without the knowledge of the nonclosure NMEs.
%, it would be difficult to pick the correct value of closure energy as there is no definite method for that. So, we also calculated
Hence, the optimal closure energy for which closure and nonclosure NMEs overlap has been calculated. The calculated optimal closure energy can be used in future calculations of the closure approach, thereby eliminating the complexity of calculating a large number of intermediate states. 
Additionally, the results of variation of NMEs for $2\nu\beta\beta$ decay of $^{124}$Sn with the excitation energy and number of states of the intermediate nucleus $^{124}$Sb are also presented. 

%In the future, it will be interesting to explore the extent to which the NMEs for $0\nu\beta\beta$ and $2\nu\beta\beta$ decay become more reliable as we include more intermediate states. 
In the future, it will be interesting to explore how the nonclosure approach can affect the NMEs for other Beyond Standard Model (BSM) mechanisms, such as the left-right symmetric mechanisms.
%===============================
\begin{acknowledgments}
We gratefully acknowledge the support provided by the Board of Research in Nuclear Sciences (BRNS), Government of India, through project grant No. 58/14/08/2020-BRNS/37085. Y. Iwata is thankful to the Tokyo Institute of Technology for generously allowing to use their high-performance computing facility to perform some nuclear states calculation using the KSHELL code.
\end{acknowledgments}
%=============================
\nocite{}
%=============================
%=============================
%=============================
\bibliography{main}% Produces the bibliography via BibTeX.
\end{document}